\documentclass[twocolumn]{revtex4}

\usepackage{amssymb}
\usepackage{amstext}
\usepackage{amsmath}

\newcommand*\de{\mathrm{d}}
\newcommand*\De{\mathrm{D}}
\renewcommand*\epsilon{\varepsilon}
\renewcommand*\phi{\varphi}
\renewcommand*\theta{\vartheta}

\begin{document}
\title{Geodesic deviation and gravitational waves} 

\author{M. Leclerc}

\affiliation{Section of Astrophysics and Astronomy, 
Department of Physics,  University of Athens, Greece}

\begin{abstract} 
The detection of gravitational waves based on the 
geodesic deviation equation is discussed. 
In particular, it is shown that the only non-vanishing 
components of the wave field in the conventional traceless-transverse 
gauge in linearized general relativity do not enter the geodesic 
deviation equation, and therefore, apparently, no effect is predicted 
by that equation in that specific gauge. 
The reason is traced back to the fact that the 
geodesic deviation equation is written in 
terms of a coordinate distance, which is not a directly 
measurable quantity. On the other hand, in the proper Lorentz frame of the 
detector, the conventional result described in standard textbooks holds. 
\end{abstract}
\maketitle

\section{Geodesic deviation} \label{geo}

We begin by giving a short derivation of  
 the equation of geodesic deviation,  
without reference to a specific theory of gravitation. It is 
only assumed that a free falling particle follows a geodesic 
with respect to a certain, unspecified, connection. 

Consider a particle with wordline $x^i(\tau)$ following a geodesic 
\begin{equation} \label{1}
\De u^i = \frac{\de u^i }{\de \tau} + \Gamma^i_{kl}(x) u^k u^l = 0, 
\end{equation}
where $u^i = \de x^i / \de \tau$. Obviously, only the symmetric 
part of the connection enters the geodesic equation, and we will 
therefore assume a symmetric connection from now on. Metric 
compatibility (or even the existence of a metric) is not necessary 
for our arguments. 

Next, consider a neighboring geodesic $\tilde x^i(\tau) = 
x^i(\tau) + \xi^i(\tau)$, following 
\begin{equation} \label{2}
\De \tilde u^i = \frac{\de \tilde u^i }{\de \tau} + \Gamma^i_{kl}(\tilde x) 
\tilde u^k \tilde u^l = 0. 
\end{equation}
From $\tilde x^i - x^i = \xi^i$, we find (the dot denoting the 
derivative with respect to the curve parameter $\tau$) 
 \begin{eqnarray*}
\ddot \xi^i &=& \dot {\tilde u}^i - \dot u^i  \\
&=& - \Gamma^i_{kl}(\tilde x) \tilde u^k \tilde u^l - 
\Gamma^i_{kl}(x) u^k u^l  \\
&=& - (\Gamma^i_{kl} + \Gamma^i_{kl,m}\xi^m) (u^k + \dot \xi^k)
(u^l + \dot \xi^l) - \Gamma^i_{kl} u^k u^l,  
\end{eqnarray*}
where we have expanded $\Gamma^i_{kl}$ around $x^i$ to first order in 
$\xi^i$. To the same order, we finally find 
\begin{equation} \label{4}
\ddot \xi^{i} = - \Gamma^i_{lm,k} u^l u^m \xi^k - 2 \Gamma^i_{kl} u^k 
\dot \xi^l. 
\end{equation}
This is the geodesic deviation equation, although still in a non-explicitly 
covariant form. We now introduce covariant derivatives 
\begin{equation}\label{5}
\De \xi^i = \dot \xi^i + \Gamma^i_{kl} \xi^k u^l, 
\end{equation}
and 
\begin{eqnarray*}
\De^2 \xi^i &=& (\dot \xi^i + \Gamma^i_{kl} \xi^k u^l)\dot{}
+ \Gamma^i_{mj} (\dot \xi^m + \Gamma^m_{kl} \xi^k u^l) u^j  \\
&=& \ddot \xi^i + \Gamma^i_{kl_,m} \xi^k u^l u^m 
+ \Gamma^i_{kl} \dot \xi^k u^l  
+\Gamma^i_{kl} \xi^k [- \Gamma^l_{pj} u^p u^j] \\ &&+ 
\Gamma^i_{mj} \dot \xi^m u^j + \Gamma^i_{mj} \Gamma^m_{kl} \xi^k u^l u^j,  
\end{eqnarray*}
where the term in $\dot u^l$ has been eliminated with the geodesic equation.  
Inserting the expression for $\ddot \xi^i$, eq. (\ref{4}), we find 
\begin{eqnarray}\label{6}
\De^2 \xi^i 
&=& \left(\Gamma^i_{lk,m}- \Gamma^i_{lm,k} + \Gamma^i_{jm}\Gamma^j_{lk}
- \Gamma^i_{jk}\Gamma^j_{lm}\right) u^m u^l \xi^k       \nonumber \\
&=&
R^i_{\ lmk} u^m u^l \xi^k. 
\end{eqnarray}

Both equations (\ref{4}) and (\ref{6}) are completely equivalent 
and can be used according to convenience. 

Let us now consider a coordinate system where $\Gamma^i_{kl} = 0$ 
on the geodesic $x^i(\tau)$  under investigation ({\it proper Lorentz
  frame}). For 
$\Gamma^i_{kl}$ to vanish along the complete geodesic, we must 
have $(\Gamma^i_{kl})\dot{} = 0$.   Then, obviously, we 
have $\De^2 \xi = \ddot \xi^i$, and (\ref{6}) reduces to 
\begin{equation}\label{7} 
\ddot \xi^i = 
\left(\Gamma^i_{lk,m}- \Gamma^i_{lm,k} \right) u^m u^l \xi^k 
= R^i_{\ lmk} u^m u^l \xi^k, 
\end{equation}
a relation that can be found in many textbooks. 
There is nothing wrong with this equation, but it can easily lead to 
misconceptions. Indeed, in the same coordinate system, we find 
from (\ref{4}) the equation 
\begin{equation} \label{8}
\ddot \xi^i =  - \Gamma^i_{lm,k} u^l u^m \xi^k, 
\end{equation}
which differs from (\ref{7}) by the term $\Gamma^i_{lk,m} u^m u^l \xi^k$. 
There is still no contradiction, since $\Gamma^i_{lk,m} u^m = 
(\Gamma^i_{lk})\dot{} = 0$, meaning that the first term in 
(\ref{7}) is zero anyway. 

We also observe that there is actually no need to 
assume that $\Gamma^i_{kl}$ 
is zero along  the complete geodesic (as long as we are interested only 
in one point), because the term containing $(\Gamma^i_{kl})\dot{}$ 
 drops out anyway from eq. (\ref{6}), as is obvious in the 
form (\ref{4}). The simple requirement $\Gamma^i_{kl} = 0$ at 
the point under investigation is sufficient to obtain the result 
(\ref{8}). Note, however, that for $(\Gamma^i_{kl})\dot{} \neq 0$, the 
form (\ref{7}) is not correct, since a further term arises from $\De^2 \xi^i$.

\section{Gravitational waves}

The previous considerations were quite general. In this section, we assume 
that $\Gamma^i_{kl}$ is the Christoffel connection of the metric $g_{ik}$ 
and the dynamics are governed by the field equations of  general relativity. 

Consider now a small perturbation 
$h_{ik}$ on a flat background, i.e., $g_{ik} = \eta_{ik} + h_{ik}$. 
In the linear approximation, general relativity leads to wave fields 
that upon imposing appropriate gauge conditions, can be assumed to 
satisfy the so-called traceless-transverse gauge conditions, i.e.,   
$h_{i0} = 0$, $h^{\alpha\beta}_{\ \ ,\beta} = 0$, $h^{\alpha}_{\ 
\alpha} = 0$ (where $i,k,\dots = 0,1,2,3$, and $\alpha,\beta,\dots = 1,2,3$). 
For the rest, 
$h_{\alpha\beta}$ is a solution to the wave equation in the flat Minkowski 
background. For details on the linearized theory, the gauge freedom and 
the wave solutions, we refer to the standard textbooks 
\cite{wald, mtw, schutz,  landau}.

Let us consider the case where initially, we have 
$u^i = (u^0, 0,0,0)$. The geodesic equation (to first order in $h_{ik}$) 
then reduces to  
$\dot u^i = \Gamma^i_{00} u^0 u^0 = (h^i_{\ 0,0} - \frac{1}{2} h_{00}^{\ \ ,i})
 u^0 u^0$. For the traceless-transverse wave, this is obviously zero, 
meaning that a particle initially at rest will remain 
at rest for  any time. This is a specific feature of the coordinate system 
related to the above gauge choice. 

Let us now proceed in the way of standard textbooks. We start from 
(\ref{7}). For $u^{\alpha} = 0$ (initially) and $\xi^{0} =0$
(as a result of an appropriate choice of the curve 
 parameters of the neighboring 
geodesics), we find, to first order in $h_{ik}$ 
\begin{equation}  \label{9}  
\ddot \xi^{\alpha} = \frac{1}{2} \ddot h^{\alpha}_{\ \beta} \xi^{\beta},  
\end{equation}
where we have used the fact that $u^0 = 1 + \mathcal O(h) $ and 
$h^{\alpha\beta}_{\ \ ,0,0} = \ddot h^{\alpha\beta} + \mathcal O(h^2)$
(in other words, to the required order, we can identify the curve parameter 
with the time coordinate).  

This is the result conventionally presented in literature. 
Similar relations have already been used by Weber in the context 
of resonant-mass wave detectors  
(see, e.g.,  eqs. (14) and (15) of \cite{weber}) 
and equation (\ref{9}) can be found in most textbooks on 
general relativity, see for instance \cite{wald}, \cite{mtw} and
\cite{schutz}. The right hand side of (\ref{9}) 
 describes the force acting  on two neighboring particles 
hold at fixed (coordinate) positions. 

At first sight, one might argue that  eq. (\ref{9}) is incorrect for the 
following reason:   Equation (\ref{7}) was derived under the assumption 
$(\Gamma^i_{kl})\dot{} = 0$, meaning that the first term in 
(\ref{7}) is  zero. But it is exactly from this term that the right hand 
side of (\ref{9}) emerges. Thus, if we insist in using eq. (\ref{7}), 
we have to conclude that $\ddot h^{\alpha\beta}= 0$. This is not the 
case for a wave solution, however. The other way around, if we 
insist that $\ddot h^{\alpha \beta} \neq 0$ (as should be the case for 
a wave solution), then obviously $(\Gamma^i_{kl})\dot{} \neq 0$, and 
thus, we cannot use the form (\ref{7}) of the geodesic deviation equation. 
In summary, eq. (\ref{7}) is not consistent with the traceless-transverse 
gauge condition. 

On the other hand, if we use eq. (\ref{8}) instead, we simply find, 
with the same conditions used above 
($u^{\alpha} = 0, u^0 =1 +\mathcal O(h), \xi^0 = 0$) 
that we have $\ddot \xi^{\alpha} = 0$, meaning that  no force is 
applied by the wave on the  particles at rest.  

Is this the correct result? Well, (\ref{8}) was  derived from (\ref{4}) 
assuming that  $\Gamma^i_{kl} = 0$ at the (spacetime) point in question, 
which is less stronger 
than its vanishing on the complete geodesic, but is still not 
suitable for the present application. To find the correct relation, it is 
secure to start directly from (\ref{4}), which holds in full generality. 
The result is 
\begin{equation} \label{10}
\ddot \xi^{\alpha} = - \dot h^{\alpha}_{\ \beta} \dot \xi^{\beta}. 
\end{equation}
In particular, for particles initially at rest (i.e., at fixed coordinate 
positions), we also have $\dot \xi^{\alpha} = 0$, and thus
\begin{equation} \label{11} 
\ddot \xi^{\alpha} = 0.  
\end{equation}
This equation seems to be in contradiction with  
equation (\ref{9}), which is claimed to hold in standard textbooks. 
Note that eq. (\ref{11}) 
also insures that $\dot \xi^{\alpha}$ remains  zero at all times, and thus, 
the equation remains valid to all times. 

\section{Discussion}

In the first version of this paper, we concluded that the standard literature 
result (\ref{9}) is incorrect and that the correct relation is given by
(\ref{11}). In the meanwhile, it was pointed out to us by experts on
gravitational waves that this conclusion is not entirely correct. Infact, 
we have misinterpreted the standard literature result, and the truth is that 
both (\ref{9}) and (\ref{11}) are correct, when interpreted in the right way. 

Let us begin with the relation (\ref{11}). It was derived directly from 
(\ref{4}), using the traceless transverse gauge for $h_{ik}$ and assuming the 
usual conditions ($u^{\alpha} = 0$, $\xi^{0} = 0$). It is therefore true that 
in the traceless transverse gauge, no coordinate acceleration is induced by the
passing wave between particles at rest. As we have pointed out already in the 
first version, this does not mean that there is no physical effect of the wave 
on the detector, because even if the coordinate distance between particles
does not change, the proper distance between them will change, and this is 
physically measurable. So far, our conclusions of the first version were
correct. 

From the apparent discrepancy between (\ref{9}) and (\ref{11}), we then
concluded that the standard result (\ref{9}) must be wrong. This, however, 
is not true. Both equations simply refer to different reference frames.
Indeed, as we have shown, eq.  (\ref{7}) holds in the proper Lorentz reference 
frame of the detector. Thus, in order to apply that equation to gravitational 
waves, we should describe the perturbation $h_{ik}$ in the same reference
frame. However, the only way $h_{ik}$ enters (\ref{7}) is via $R^i_{\ lmk}$, 
which is gauge invariant. Thus, in order to evaluate $R^i_{\ lmk}$, we can 
use a gauge of our choice, the result will always be the same. The result is of
course eq. (\ref{9}). 

The important thing here is that, although we have used the traceless
transverse gauge to evaluate $R^i_{\ klm}$, the final relation (\ref{9}) is 
valid only in the proper Lorentz reference frame of the detector. 
It is therefore not in contradiction with  the result (\ref{11}), which holds
only in the traceless transverse gauge. The confusion originates from the fact 
that in eq. (\ref{9}), we have expressed $R^i_{\ klm}$ with the help of 
$h_{ik}$ in the traceless transverse gauge, which is not the  metric 
perturbation in the reference frame the equation actually holds. In other 
words, in the proper Lorentz reference frame, the metric in not given by 
$g_{ik} = \eta_{ik} + h_{ik}$.   

This might appear trivial a posteriori, but judging from several referee
reports of well established journals, which contained various confusing 
arguments against publication of the first version of the paper, 
but which did not uncover the actual reason 
for our erroneous conclusions, it seems that we are not the only ones to 
which the situation was not entirely clear. For that reason, we wrote this 
second version of the paper rather than to simply withdraw the original 
preprint. Further, an article has been pointed  
out to us \cite{bor}
which seems to contain conclusions similar to those of our original 
paper, namely that there is no effect of a gravitational wave on neighboring 
particles if they are initially at rest. Again, this conclusion 
should be related to the use of a specific reference frame.

\end{document}